\documentclass[11pt,twoside]{book}
\usepackage{vshalo}
\usepackage{longtable}
\usepackage{amsmath,amssymb}
\usepackage{graphicx}
\usepackage{lscape}
\usepackage{psfig}
\usepackage{epsf}
\usepackage{index}
\usepackage{natbib}
\usepackage{bigdelim}
\usepackage{multirow}
\makeindex
\newindex{aut}{adx}{and}{\Large Author  Index}
\newcommand{\axindex}[1]{\index[aut]{#1}}

\begin{document}

%%%%%%%%%%%%%%%%%%%%%%%%%%%%%%%%%%%
\pagestyle{myheadings}
\setcounter{equation}{0}\setcounter{figure}{0}\setcounter{footnote}{0}\setcounter{section}{0}\setcounter{table}{0} \setcounter{page}{63}

\markboth{Chuprikov \&  Guirin}{SS433: Evolution of Radio Structure in $L$ and $C$ Frequency Ranges}
\title{SS433: Evolution of Radio Structure in $L$ and $C$ Frequency Ranges}
\author{A. A. Chuprikov \& I. A. Guirin}\axindex{Chuprikov, A. A.}\axindex{Guirin, I. A.}
\affil{Astro Space Center of P.N. Lebedev Physical Institute of Russian Academy of Sciences, Moscow, Russia}

\begin{abstract}
We analyze results of data processing of observations of well-known object titled SS433 with the and 
VLBA during more than 10 years. Data have been processed with the software project titled {\it 'Astro Space 
Locator' (ASL for Windows)}. The Multi Frequency Synthesis (MFS) method has been used for reconstruction 
of radio maps at 18 and 6 centimeter wavelength ranges (L and C ranges). High quality images of SS433 for 
several epochs are presented. Evolution of its radio structure is demonstrated. Astrophysical parameters of 
object and their changes in time are discussed. Any polarization phenomena are not taken into account. We 
present results of processing of data of RR polarization for all the observational sessions
\end{abstract}

\section{Introduction}
The object titled {\it SS433 (J1911+0458)} is known more than 30 years. It associated with the Super Nova Remnant 
titled {\sl W50} which is approximately 10000 years old. SS433 is very weak in optic (the visible magnitude value 
is about +14). On the other hand, object is a powerful source of radio and X-ray emission. This close binary system 
concists of the A-class star and some dark component. Distance to SS433 is approximately 5.5 kpc 
(see Brundell(2004)). Thus, 1 milliarcsecond in map corresponds to the linear size of approximately 1 Astronomical 
Unit (1 AU = $ 1.5 \cdot 10^{8} $ km). There are the following reasons of our interest to this object :\\

$\bullet$ SS433 is included into source list of the scientific program of the 'Radioastron' ground-space 
VLBI mission (see Hirabayashi(2000)). Thus, it is necessary to have detailed information about the radio 
structure of this object and its variability

$\bullet$ object is sufficienly weak in L and C frequency ranges. It is relevant for advancing of 
calibration procedures implemented into our software project

$\bullet$ SS433 has been observed with VLBA many times for different epochs\\

In this paper we analyze results of processing of data of the following observational sessions : {\it GP025, 
GP050, BC174, BM111}. All the data were tranferred from archive of the {\it NRAO (National Radio Astronomy Observatory, 
USA)} and processed with with the software project titled 'Astro Space Locator' (ASL for Windows) (see Chuprikov(2002)).

\section{Method and Results of Data Processing}
The data processing consists of the following stages :\\

$\bullet$ Amplitude calibration of all the data using values from GC and TY tables and some additional
information

$\bullet$ Single band fringe fitting (the primary phase calibration) of all the data. 
Estimation of the optimum value of solution interval

$\bullet$ Averaging of all the data over each frequency band

$\bullet$ Multi band fringe fitting of the atmosphere and ionosphere calibrator data. Estimation of gain 
values for every frequency and every time interval

$\bullet$ Application of gains, compensation of atmosphere and ionosphere delay for all the data

$\bullet$ Self-calibration. Final averaging, editing of the data, and {\sl Imaging}\\

Amplitude calibration of the data was made with usage of two standard calibration tables 
({\sl 'Gain Curve' (GC)} and {\sl 'System Temperature' (TY)}) that were established 
with the VLBA correlator during the primary data processing. The standard procedure of primary phase calibration 
allows to reconstruct the visibility function phase. Then, we can obtain the dirty map of our source and estimate 
the signal-to-noise ratio (SNR). For weak sources, such as SS433, SNR depends conciderably on the phase calibration 
solution interval value. It is necessary, to find the optimal solution interval for every particular data set. In 
our case, this value changes between 10 seconds and 8 minutes. After phase of visibility function is 
reconstructed, all the data could be averaged in time and frequency. SNR will 
be increased due to this averaging. The main difficult stage of the VLBI data processing is usage of phase calibrators 
for compensation of interferometer model errors as well as for compensation of atmosphere and ionosphere delays. For 
this goal, it is necessary to include into any VLBI session schedule some additional sources (phase calibrators). 
There are two criteria for selection of such sources :\\

$\bullet$ They are close to the main source (at the distance no more than some degrees)

$\bullet$ They are point-like and bright enough\\

In our case, the sources titled {\it J1913+0508, J1929+0507, J1907+0127} and {\it J1950+0807} were 
used as such calibrators. As mentioned above, they are used to estimate the gain values for compensation of 
delays caused by atmosphere, ionosphere and any uncertainties of the interferometer model.

It is well known (see Doolin (2009)) that SS433 has at least 4 periodical modes :\\

$\bullet$ Orbital period is approximately 13.08 days

$\bullet$ Jet traces a cone with angle of approximately 40 degrees with period of 162.375 days

$\bullet$ There is a nodding superimposed in the precession of the jet axis with a period of 6.06 days

$\bullet$ There is a prcession of ruff independently on the jet precession. This period is equal to 552.5 days\\

Thus, a radio structure of object is very changeble in time. Figure 1 demonstrates changes in radio structure of 
SS433 in C range (6 cm). The ruff is seen clearly in the left picture. Its size is about 10 mas (or 10 AU). 
The length of visible part of jet is approximately 30 mas (30 AU). The right picture shows the same object at 
other epoch. Here, the size of ruff seems to be conciderably less (about 5 - 10 AU). Moreover, it is clear 
that position angles of jet are strongly different for these two epochs (difference is about 20 degrees) and 
length of jet in the right picture is about 1.5 times less than in the left picture.
 
\begin{figure}[!h]
\centerline{\hbox{\psfig{figure=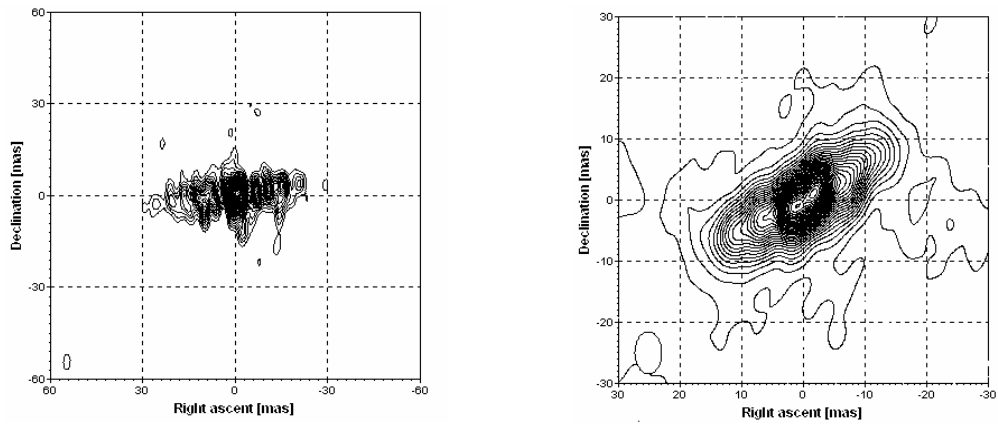,angle=0,clip=,width=13.3cm}}}
\caption[]{The 6 cm radio images reconstructed of SS433. Session {\it BM111}, 18/11/1998 (LEFT).
Session {\it BC174}, 8/5/2008 (RIGHT)} 
\label{sterken-fig1} 
\end{figure}

Figure 2 demonstrates changes in radio structure of SS433 in other wavelength range (18 cm, L range). 
It is clear, that object is much more changeble in this range. The length of visible part of jet is here 
about 200 mas (200 AU). Ruff is not seen so clearly in this frequency range. It seems to be visible in the 
right picture. Size of ruff seems to be about 20 - 40 AU. Again, it is clear that position angles of jet are 
strongly different for these two epochs (difference is about 40 degrees). In the right picture, length of jet 
is about 4 times less than in the left picture.

\begin{figure}[!h]
\centerline{\hbox{\psfig{figure=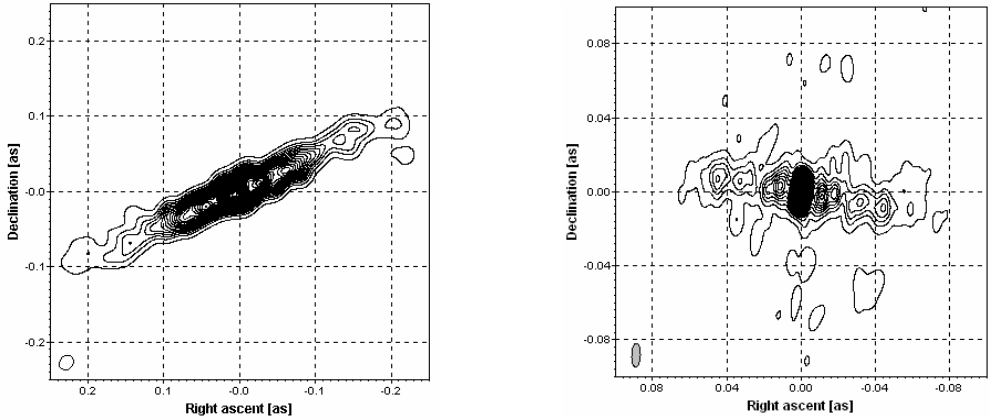,angle=0,clip=,width=13.3cm}}}
\caption[]{The 18 cm radio images reconstructed of SS433. Session {\it GP050}, 30/01/1999 (LEFT).
Session {\it GP025}, 20/02/2000 (RIGHT)} 
\label{sterken-fig2} 
\end{figure}

\section{Conclusions}
We made processing of data of some VLBA observational sessions had been made during 1998 - 2008 
{\it GP025, GP050, BC174, BM111}. The main goal of this investigation is to reveal the radio structure 
of SS433 and its changing in time. Results of these data processing are :\\

$\bullet$ The radio images reconstucted in both L (18 cm) and C (6 cm) wavelength ranges confirms the existence 
of precessing ruff in radio structure of the object. The size of the ruff is approximately 40 AU for the L range. 
It more compact for the C range (less than 20 AU for the C range)

$\bullet$ This investigation demonstrates that calibration procedures of 'Astro Space Locator' are relevant 
for processing of interferometrical data for weak radio sources\\

We'll continue our investigations to reveal more detail information of variability of radio structure 
of SS433 during one orbital period as well as the modulation of this variability.

\section{Acknowlegements}
The authors would like to thank Dr. Vitaly Goranskij (Sternberg Astronomical Institute, Moscow University, 
Russia) for very useful discussion about the nature of the SS433.

\end{document}